\documentclass[12pt]{iopart}
\usepackage{amssymb,amsfonts,bbm,graphicx,sidecap,epstopdf}
\usepackage{braket,color}

\newcommand{\unit}[1]{\ensuremath{\, \mathrm{#1}}}
\newcommand{\eqref}[1]{(\ref{#1})}

\def\d{{\rm d}}
\def\e{{\rm e}}
\def\ii{{\rm i}}

\def\k{\kappa}

\def\positron{$e^{+}$~{}}
\def\antiproton{$\overline{p}~{}$}
\def\positronium{Ps~{}}

\newcommand{\STErev}[1]{#1}

\begin{document}
\title[Matter-wave interferometry: towards antimatter interferometers]{Matter-wave interferometry:\\ towards antimatter interferometers}
\author{Simone Sala$^{1,2}$, Fabrizio Castelli$^{1,2}$, Marco Giammarchi$^2$,\\
Stefano Siccardi$^1$ and Stefano Olivares$^{1,2}$}
\address{$^1$ Dipartimento di Fisica, Universit\`a degli Studi di Milano,
I-20133 Milano, Italy}
\address{$^2$ INFN Sezione di Milano, I-20133 Milano, Italy}
\ead{stefano.olivares@fisica.unimi.it}

\date{\today}
\begin{abstract}
\STErev{Starting from an elementary model and refining it to take into account more realistic effects, we discuss the limitations and advantages of matter-wave interferometry in different configurations. 
We focus on the possibility to apply this approach to scenarios involving antimatter, such as positrons and positronium atoms. In particular, we investigate the Talbot-Lau interferometer with material gratings and discuss in details the results in view of the possible experimental verification.}
\end{abstract}

\section{Introduction}\label{s:intro}

Matter-wave interference is at the heart of the quantum mechanical
nature of particles. While this phenomenon has been observed for electrons \cite{merli,merli:2,tonomura}, neutrons \cite{neut:88,rauch}, atoms and molecules \cite{arndt,arndt2,RMP:09} using a variety of different experimental tools, no \STErev{experimental} tests exist on elementary antimatter particles, or
matter-antimatter systems. However, beams of antiparticles at low energy are becoming increasingly
available, as in the case of antiprotons at the CERN Antiproton Decelerator \cite{AntiprotonDecelerator} or in the case of positrons (and the associated positronium production) in $^{22}$Na source systems coupled with Surko traps \cite{surko}.
\par
\STErev{In this paper we discuss the optical analogy and the main principles of
Fraunhofer and Talbot matter-wave interference regimes, considering
material gratings, in order to introduce the issues and the
problems of antimatter interferometry.} Positrons ($e^{+}$) are proposed as our first antimatter system to study and positronium (Ps) is the atom that we will be considering as a matter-antimatter symmetric system. The antiproton ($\overline{p}$) case will also be shortly discussed.
\par
The paper is structured as follows. In section \ref{s:basic} we review the basic elements of quantum diffraction theory of particles from a grating and describe the build-up of the statistical interference pattern. Section \ref{s:advanced} focuses on the incoherence due to the source, such as the effect of the particle velocity spectrum and the source geometrical extension. In section \ref{s:interactions} we address the interaction between particles and a grating considering both neutral particles and charged particles. Section \ref{s:talbot} is devoted to Talbot-Lau interferometry: we describe the geometry, and its advantages with respect to single grating setups.
Furthermore, we numerically show how the fringe visibility is affected by the particle velocity spread, when realistic parameters are used to carry out Monte Carlo simulated experiments. Finally, we close the paper drawing some concluding remarks in section \ref{s:conclusion}.

\section{Basic quantum model of diffraction}\label{s:basic}

\begin{figure}[tb]
\centering
 \includegraphics[width=0.3\textwidth,keepaspectratio]{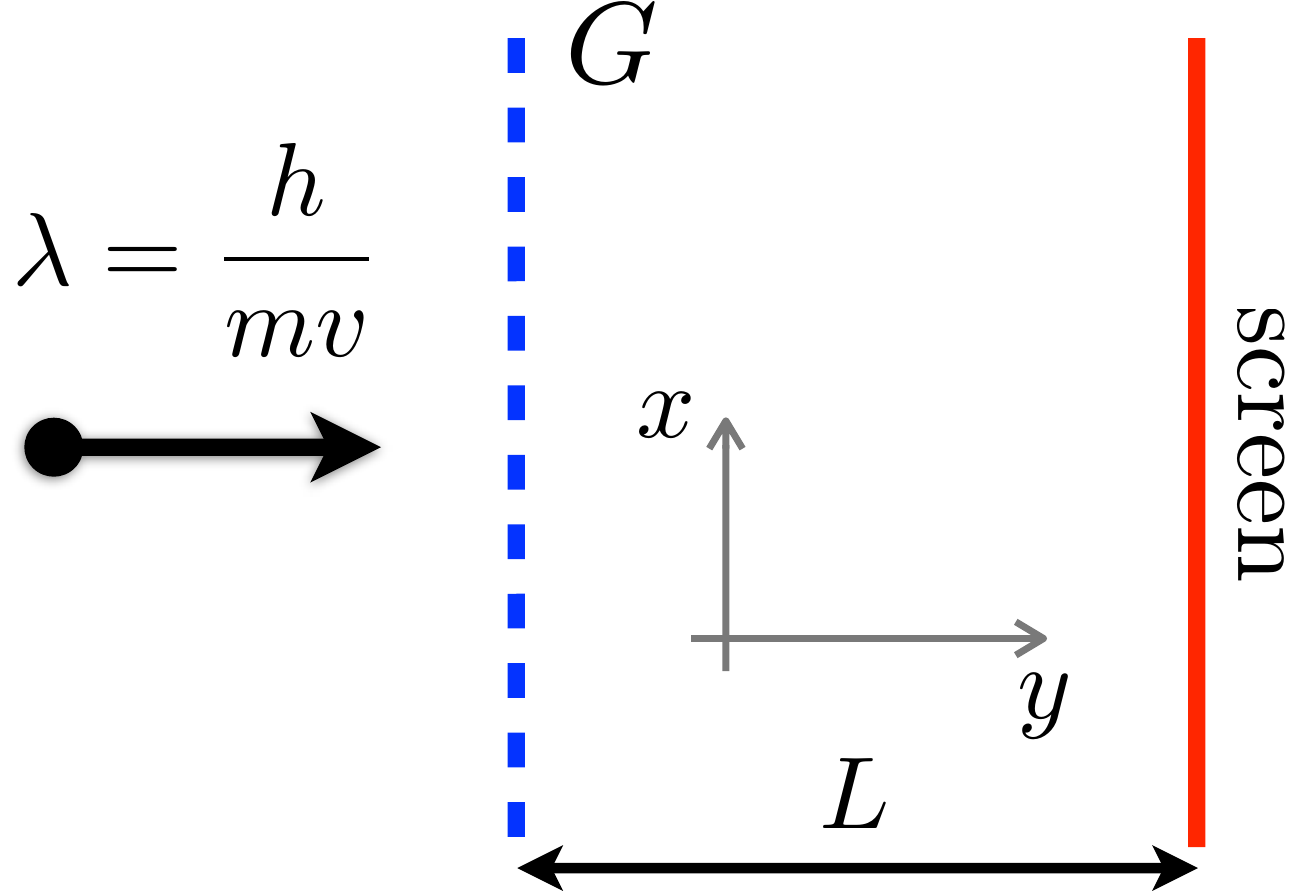}
  \caption{(Color online) A single particle of momentum $p=mv$ impinging on an $N$-slit grating ($G$); detection will take place on a screen, placed at a distance ($L$). The grating has a period $D$ and the width of each slit is $a$. The $z$-axis is orthogonal to the $x$--$y$ plane.}
  \label{fig:grating}
\end{figure}

In this section we review the basics of matter-wave interferometry. We assume that a particle moving along the $y$-axis with de Broglie wavelength $\lambda = h / (m v)$, $m$ and $v$ being respectively its mass and its velocity along the $y$ axis, interacts with an $N$-slit grating laying in the $x$--$z$ plane (see Fig.~\ref{fig:grating}). Upon assuming that the slits are sufficiently large along the $z$-axis, so that diffraction is negligible along that direction, we can represent the state just after the grating at time $t=0$ and $y=0$, being $t=y/v$, as the following superposition state \cite{facchi,vialezanghi,sanzborondo}:
\begin{equation}
\psi^{(N)} (x,t=0) \propto \sum_{n=1}^N \psi_n(x,t=0),
\label{eqn:superposition}
\end{equation}
where $\psi_n(x,t=0)$, $n=1,\ldots , N$, are the wave functions describing the particle passed through the $n$-th slit (we assume, as usual, that the slits are independent). For a system of identical slits with period $D$, we can write $\psi_n(x,0) =  \psi_0(x-nD)$. Indeed, the actual expression of $\psi_n (x,t=0)$ is dictated by the characteristics of the diffraction grating and its interaction with the incoming particle. As the grating prepares the system in the state of Eq.~(\ref{eqn:superposition}), we can assume that the motion along the $x$-axis is governed by the free Hamiltonian:
\begin{equation}
H_{\rm{eff}}=\frac{p_x^2}{2m}.
\label{eqn:hamiltonian} 
\end{equation}
Therefore, the evolved state $\psi(x,t)$ is obtained by solving the Schr{\"o}dinger equation with the Hamiltonian (\ref{eqn:hamiltonian}). In particular, the particle probability density distribution along the $x$-axis on the screen at position $y=L$ (the interference pattern), is given by $I(x) = |\psi^{(N)}(x,t=L/v)|^2$, where: 
\begin{equation}\label{class:opt}
\psi^{(N)}(x,t) =  \frac{1}{\sqrt{\lambda L}} \int_{-\infty}^{+\infty} 
\exp\left[\ii \frac{ \pi}{\lambda L} (x-x')^2\right]\, \psi^{(N)}(x',0)	\, \d x' ,
\end{equation}
that is formally identical to the Fresnel integral of classical optics \cite{goodman}.
The most common approach found in literature \cite{vialezanghi,sanzborondo,olivares1,olivares:sa} is to adopt an ``effective'' point of view and postulate a convenient form for the initial single-slit wave function, for example:
\begin{equation}
\psi_n(x,0)= a^{-1} \chi_{[-\frac{a}{2}+nD,\frac{a}{2}+nD]}(x)
\label{eqn:rect}
\end{equation}
where $a$ is the slit width and
$$
\chi_{\Omega}(x)=
\left\{
\begin{array}{ll} 1 & \mbox{if}\, x\in \Omega, \\[1ex]
0 & \mbox{otherwise}.
\end{array}
\right.$$
This is the quantum mechanical analog of assuming uniform illumination in the treatment of light diffraction. Another useful choice for the initial single-slit wave function is a Gaussian function centered on the slit interval with a suitable variance $\sigma$, namely (we drop the overall normalization constants);
\begin{equation}
\psi_n(x,0)=\exp\left[-\frac{(x - nD)^2}{4 \sigma ^2}\right].
\label{eqn:gaussian}
\end{equation}
This choice is more convenient, as many calculations can be easily carried out analytically on Gaussian functions. In this case, the parameter $\sigma$ is usually set to $\sigma=a/(2 \sqrt{2 \pi})$.
Upon introducing the rescaled variables: 
\begin{equation}
\hat{x} = \frac{x}{\sigma}, \quad \hat{D} = \frac{D}{\sigma}, \quad
\mbox{and}\quad \hat{L} = \frac{\hbar t}{2 m \sigma^2} =\frac{L \lambda}{4 \pi \sigma^2},
\label{eqn:adim}
\end{equation}
and considering a two-slit setup, the time evolved wave function outgoing from a double slit setup reads
\begin{equation*}
\psi_2(\hat{x},\hat{L})= \sum_{n=1,2} C_n \exp \left[ -\frac{\left( \hat{x} - \hat{x}_n \right)^2}{4 \left( 1 + \hat{L}^2 \right)} \left( 1 - \ii \hat{L} \right) \right]
\end{equation*}
\noindent where $\hat{x}_1 = -\hat{D}/2$ and $\hat{x}_2 = +\hat{D}/2$. The generalization for a set of $N$ equally separated slits is straightforward.
We have introduced the relative normalization constants $C_n$, to account for a possible asymmetry in the beam preparation \cite{sanzborondo}. In the following we assume perfect symmetry  $C_1 = C_2 =1$. After simple algebraic manipulations, defining
\begin{equation*}
F_{\pm} = \exp \left[ -\frac{(\hat{x}\pm \hat{D}/2)^2}{2(1 +\hat{L}^2)} \right]
\end{equation*}
\noindent the intensity reads
\begin{equation}
I(\hat{x},\hat{L}) = F_{+} +F_{-} + 2 \sqrt{F_{+} F_{-}} \cos \left[ \frac{\hat{L} \hat{x} \hat{D}}{2(1+\hat{L}^2)} \right]
\label{eqn:intensity2}
\end{equation}
\begin{figure}[htbp]
\begin{center}
    \includegraphics[width=.6\textwidth,keepaspectratio]{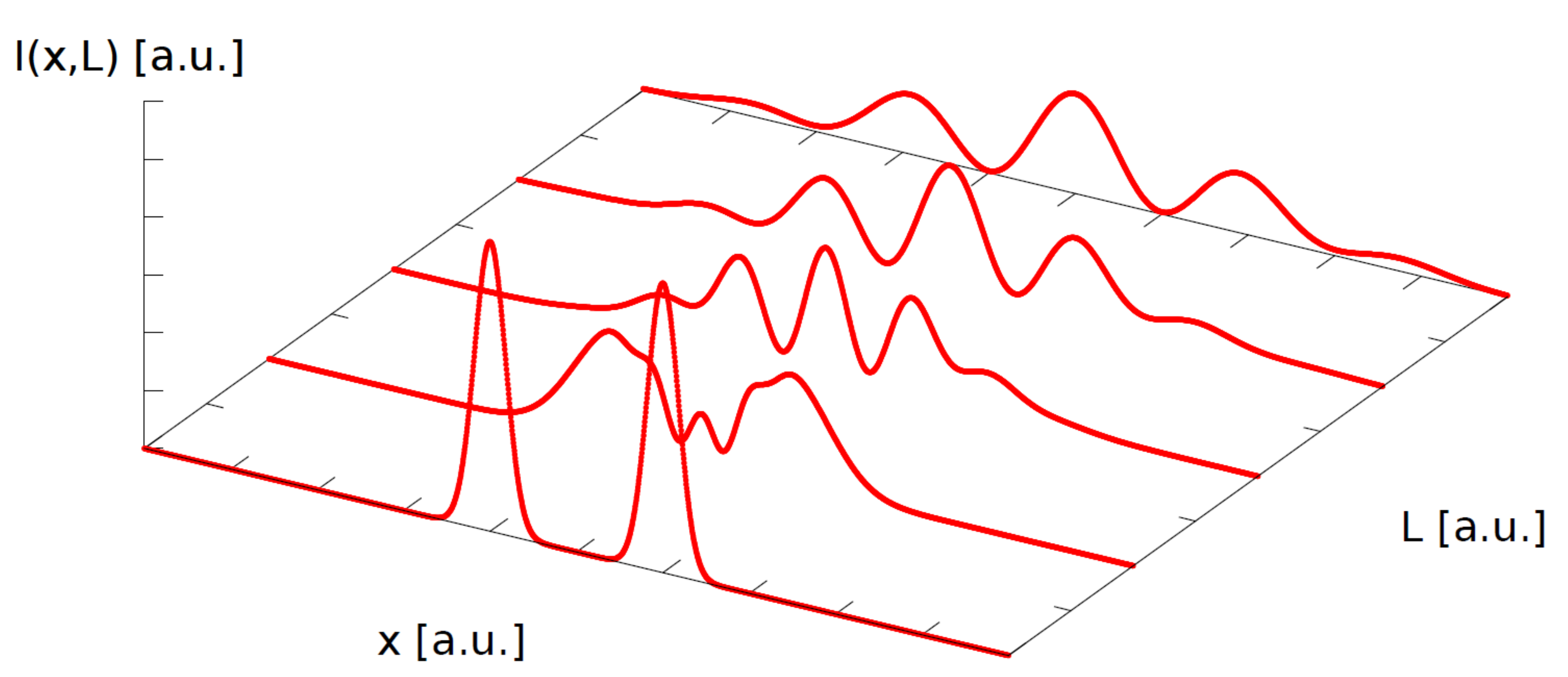}
  \end{center}
  \vspace{-0.5cm}
  \caption{\label{f:int:pattern}\STErev{The evolution of the interference pattern \eqref{eqn:intensity2} shown with its dependence on the screen distance $L$.} }
  \label{fig:dualgauss3}
\end{figure}
which clearly shows the appearance of an interference pattern due to the oscillating term (see Fig.~\ref{f:int:pattern}). It is worth noting that the condition for observing the interference maxima in far field turns out to be the usual relation of classical optics, thus in the limit $\hat{L} \gg 1$, the condition for observing a maximum reduces to
\begin{equation*}
\frac{\hat{x} \hat{D}}{2 \hat{L}} = 2 n \pi \rightarrow \frac{x D \pi}{ L \lambda} = n \pi
\end{equation*}
which is indeed the expected classical relation. The formal analogy with classical optics [see Eq.~(\ref{class:opt})] also ensures that the choice of the initial single slit profile impacts only the envelope of the intensity pattern and not its oscillatory behavior.
The classical Fraunhofer field outgoing a double slit setup reads
\begin{equation*}
I_{\rm{class}} (x,L) \propto \mbox{sinc}^2\left(\pi a \frac{xL}{\lambda}\right) \left[ 1 + \cos \left( 2 \pi D \frac{xL}{\lambda} \right) \right]
\end{equation*}
while starting from \eqref{eqn:intensity2} it easy to recover a Fraunhofer-like expression by taking the far field limit in the form $\hat{L} \gg 1 $ and $\hat{x} \gg \hat{D}$, so that 
\begin{equation*}
F_{+} =  F_{-} \simeq \exp \left( -\frac{\hat{x}^2}{2\hat{L}^2} \right)
\end{equation*}
and finally, in order to highlight the similarity with the classical expression in the Fraunhofer limit, we use Eqs.~\eqref{eqn:adim} in Eq.~(\ref{eqn:intensity2}), obtaining:
\begin{equation}
I(x,L) = 2 \exp \left[ -2\left( 2\pi\, \sigma\,
\frac{x}{ \lambda  L} \right)^2 \right]
\left[ 1 + \cos \left( 2 \pi D \frac{xL}{\lambda} \right) \right].
\label{eqn:intensity23}
\end{equation}
So, much alike the classical case, in a quantum treatment based on the free evolution of single-slit wave functions, the latter factorizes and determines the envelope of the pattern.

\section{Incoherence due to the source}\label{s:advanced}

In order to describe a real experiment, the model introduced so far is not enough, since many relevant departures from the ideal situation arise. For instance, the particles' speeds vary according to a given distribution, the particle source has a finite size and the collimation stage unavoidably introduces transverse momenta. Furthermore, focusing on the scenario we are interested in, unstable antimatter atoms like \positronium can decay in flight.
All of these issues lead to \emph{incoherence} effects. 
In general, if ${\bf q} = (q_1,q_2,\ldots)$ is the vector of the physical parameters $q_k$ which can classically fluctuate in a real experiment, we can describe the overall incoherence effect by averaging the ideal intensity $I(x,t|{\bf q})$ given a suitable distribution $p({\bf q})$, that is:
\begin{equation}
\bar{I}(x,t) = \int I(x,t|{\bf q}) \, p({\bf q})\, \d{\bf q}
\label{eqn:mcmodel}
\end{equation}
There are two relevant examples of incoherence: the one due to  a finite transverse coherence length, the other due to the presence of a non-monochromatic beam. In this section we focus on the first one, whereas the effects of a non-monochromatic beam will be considered in section~\ref{s:talbot} in the context of the Talbot-Lau interferometry.
\begin{figure}
\centering
        \includegraphics[width=0.4\textwidth]{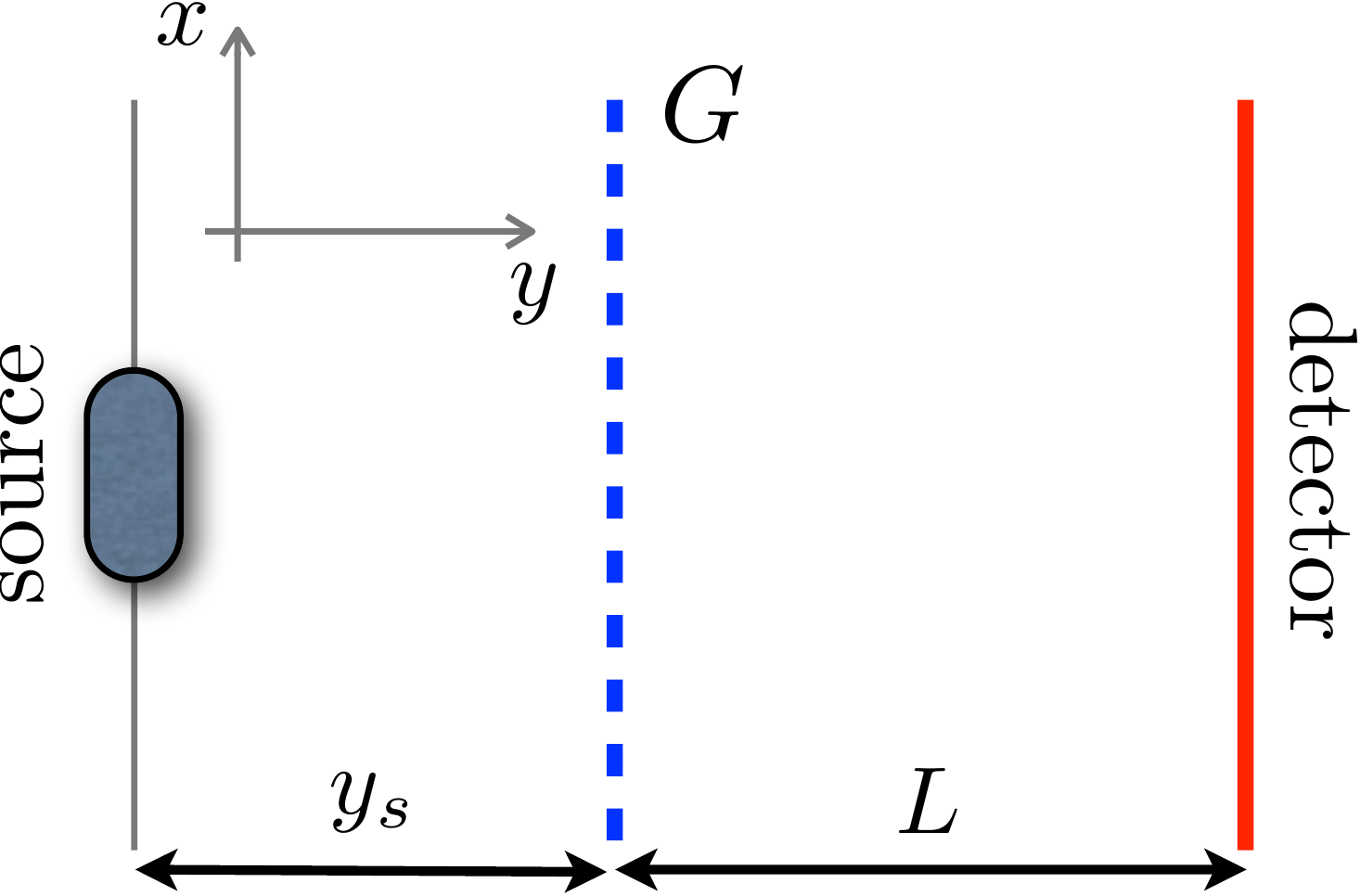}
  \caption{Sketch of an interferometer operating in the far field, where an incoherent extended source (transverse size $\sigma_s$) illuminates an $N$-slit grating G (period $D $ and slit width $a$) from a distance $y_s$.}
  \label{fig:gratingwithsource}
\end{figure}

The experimental results show that the patterns of matter-wave experiments with multi-slit gratings can be described by considering a limited number of slits \cite{olivares1,vialefull}. We define an experimental parameter $l_0$, the \emph{coherence length}, as the typical transverse length scale on the plane of the grating that sets how many slits can coherently take part to the interference process. From the physical point of view, a finite coherence length is a result of both the spatial extension and the intrinsic incoherence of the sources typically employed in matter-wave experiments. In order to take into account this effect, we add a (common) random transverse momentum $k_x$ along $x$-direction to the wave function $\psi_n(x,t=0)$ associated with each slit, namely, $\psi_n(x,t=0)\,\exp\left(\ii xk_x\right)$. If we assume that $k_x$ is distributed according to a Gaussian distribution with zero mean and variance $1/l_0^2$, one obtains the following analytical result for the intensity, valid in the Fraunhofer limit \cite{vialefull}:
\begin{eqnarray}
\fl I'(x,L) = \frac{2 \pi N}{\lambda L} \left|\hat{\psi}(x)\right|^2
\left\{ 1+  2\sum_{n=1}^{N-1}\frac{N-n}{N}\exp\left[-\frac{(nD)^2}{2 l_0^2}\right] \cos \left( \frac{2 \pi n D x }{\lambda L}\right)\right\},
\label{eqn:vialezanghi}
\end{eqnarray}
where
\begin{equation}\label{envelope}
\hat{\psi}(x) = \int\psi(x')\,\exp\left(\ii\frac{2\pi x}{\lambda L}\,x'\right) \, \d x',
\end{equation}
is an envelope corresponding to the rescaled Fourier transform of the single-slit wave function $\psi(x)=\psi_0(x,t=0)$. Since for $nD \gg l_0$, the corresponding exponential term suppresses the interference, $l_0$ can be regarded as the coherence length predicted in Ref.~\cite{vialefull}, which is inversely proportional to the transverse momentum spread. This could allow to determine the coherence length {\it a priori}, without resorting to a fit of the model to experimental data. However the transverse momentum distribution is not always easily guessed. For example, if two successive slits are used as collimators, a bound on the maximum transverse momentum could be established with a geometrical construction \cite{vialefull}. Nevertheless, since we are interested in more compact geometries as in the presence of in-flight decay of unstable antimatter, limiting the dimensions of the apparatus will be of the utmost importance. So we would need an estimate on the coherence length for an apparatus of the kind of Fig.~\ref{fig:gratingwithsource}.
\par
We can obtain the intensity within the framework of the model given  in Eq.~(\ref{eqn:mcmodel}) as follows. We assume that at random time a particle is emitted with a speed $v$ from a point $x_s$ of the source (located at the distance $y_s$ from the grating), following a distribution $p(x_s,v)$ that is determined by the nature of the source itself. After its emission the particle crosses the grating and produces an interference intensity pattern that depends parametrically on these quantities. Under the same assumption there discussed, the overall intensity at the screen is thus given by Eq.~(\ref{eqn:mcmodel}), that now reads:
\begin{equation}\label{I:bar}
\bar{I}(x,L) = \int I(x,L | x_s , v)\, p(x_s,v)\,  \d x_s\, \d v.
\end{equation}
The integration can be performed via Monte Carlo (MC) method, as it scales well with the dimension of the parameter space. Moreover, we can refine our analysis, e.g., taking into account the instability of the particles and their lifetime. As a first approximation, we could simply discard the particles that do not reach the detector plane. This corresponds to employ a detector able to discriminate between a true event and the background noise induced by the decay in flight. 
\par
\begin{figure}
\centering
    \includegraphics[width=0.5\textwidth]{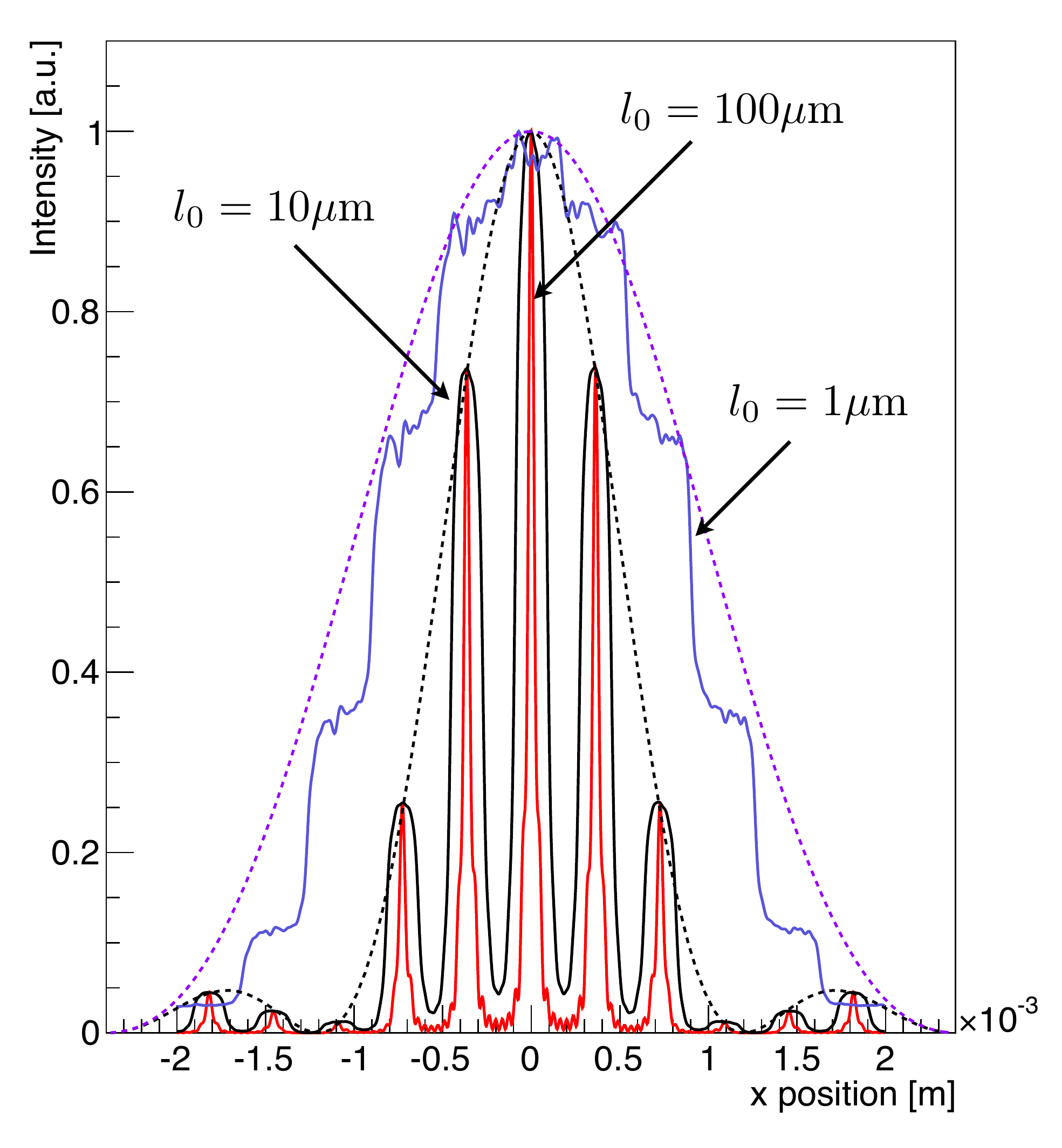}
        \vspace{-0.4cm}
\caption{\label{f:contrast} \STErev{Monte Carlo simulation of the far field setup (Fig.~\ref{fig:gratingwithsource}) from Eq.~(\ref{I:bar}) for a monochromatic beam of Ps atoms with $\lambda = 3.6 \unit{nm}$ ($m_{\rm Ps} = 2 \, m_{e}$, $v = 10^5 \unit{m/s}$) and different values of the 
source dimension $\sigma_s= 900, 90$ and $9  \unit{\mu m}$, corresponding to the
coherence length $l_0= 1, 10$ and $100 \unit{\mu m}$ (as shown in the plot). We also set $D=10 \unit{\mu m}$, $a=3 \unit{\mu m}$, $N=10$, $y_s=0.5 \unit{m}$ and $L=1 \unit{m}$. Interference disappears when $l_0 < D$, and the contrast starts to decrease when $l_0 \approx D$. The dashed curves refers to the corresponding single-slit diffraction envelopes.}}
\end{figure}

In order to obtain the same results as in Eq.~\eqref{eqn:vialezanghi} for a suitable choice of $l_0$ and in the Fraunhofer approximation, we should consider a monochromatic beam and average only over the source dimension $x_s$, assuming a \emph{uniform} probability density $p(x_s)=\sigma_s^{-1}\,\chi_{[-\sigma_s/2,\sigma_s/2]}(x)$, $\sigma_s$ being the source dimension. It is worth noting that in our simulations the average intensities are computed retaining the full accuracy of the Fresnel integral, i.e., without the Fraunhofer approximation. In a setup like the one shown in Fig.~\ref{fig:gratingwithsource} a comparison between Eqs.~(\ref{eqn:vialezanghi}) and (\ref{I:bar}) shows that the coherence length $l_0$ can be estimated as  \cite{atomintgeneric}:
\begin{equation}
l_0 \approx \frac{y_s \lambda}{2\sigma_s}.
\label{eqn:cohestimate}
\end{equation}
The dependence on the physical parameters is in agreement with the naive estimate associating the coherent illumination region in this kind of setup with the width of the central diffraction peak for a slit of size $\sigma_s$, where $\sigma_s$ is the transverse extension of the source \cite{atomintgeneric}.
\STErev{In Fig.~\ref{f:contrast} we show the Monte Carlo simulations of the far field interference pattern for different values of the source dimension $\sigma_s$ and a particular choice of the other involved parameters. For the simulations we considered a typical Ps velocity $v = 10^5 \unit{m/s}$  \cite{positronium:3}, which leads to $\lambda = 3.6 \unit{nm}$.} 
We note that as the coherence length approaches the critical value $D$, we observe a decrease in the contrast or visibility of the pattern (see Fig.~\ref{f:contrast}):
\begin{equation}
C = \frac{I_{\rm max} - I_{\rm min} }{I_{\rm max} + I_{\rm min}}
\end{equation}
defined as the difference between the intensity of a maximum and its adjacent minimum.
We will reconsider the implication for the design of an experiment in sect.~\ref{s:talbot}.

\section{Interaction with material gratings}\label{s:interactions}

In the previous section we addressed the interferometry problem assuming that the particle did not interact with the grating. As physical quantum mechanical objects the particles interact in various ways with the walls of the material grating. The formalism we developed so far is sufficiently general to account for this effect by modifying the initial wave function accordingly.
\par
\begin{figure}[tb]
	\centering
    \includegraphics[width=0.35\textwidth]{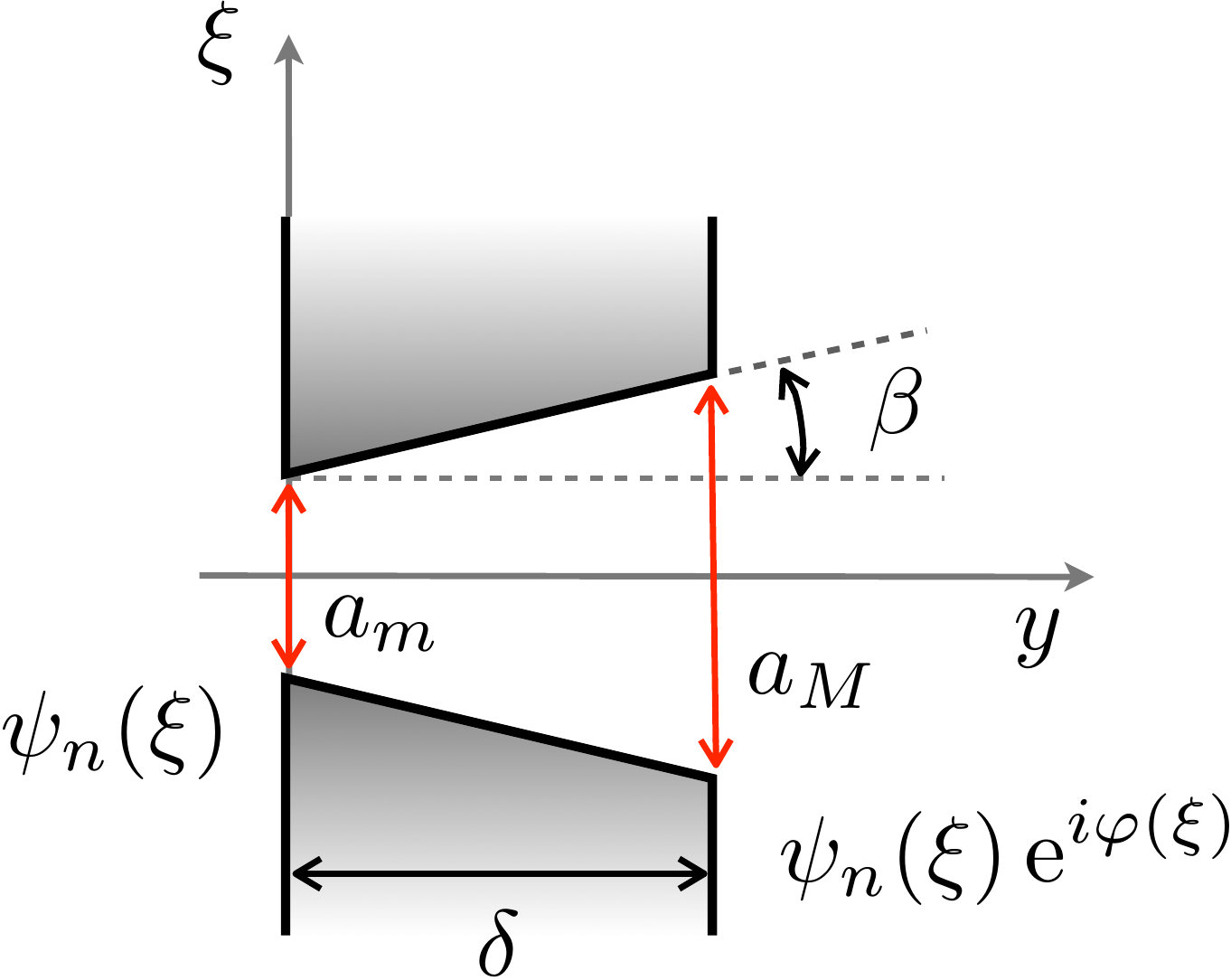}
    	\vspace{-0.3cm}
  \caption{Trapezoidal bars cross section with the narrower side of the slit facing the beam, displayed for a wedge angle $\beta$ and nominal maximum and minimum widths $a_M$ and $a_m=a_M - 2 \delta \tan \beta$, respectively.}
  \label{fig:wedgeslit}
\end{figure}

In the previous sections we have seen that the fundamental building block for quantum models of diffraction from a grating is the single slit outgoing wave function $\psi_n(\xi,0)$, which is usually postulated to be of either Gaussian or rectangular shape. If the potential $V(\xi,y)$ acting on the particle is known in the region within one slit, we can account for the interaction  by treating the slit as a \emph{phase mask}, producing a transmission function of the form $t_A(\xi)=e^{\ii \varphi(\xi)}$ \cite{vdw,gris:02,savas}, implying the substitution (see Fig.~\ref{fig:wedgeslit})
\begin{equation}
\psi_n(\xi,0) \to \psi_n(\xi,0)\, \e^{\ii \varphi(\xi)} .
\label{eqn:outputwavefunction}
\end{equation}

The standard approach \cite{vdw,savas,cornu} is now to determine the phase shift $\varphi(\xi)$ via the semiclassical eikonal approximation. Denoting with $v$ the particle speed we can write:
\begin{equation}
\varphi(\xi) = - \frac{1}{\hbar v} \int  V(\xi,y)\, \d y.
\label{eqn:vdwphaseshift}
\end{equation}
As discussed, the Fourier transform $\hat{\psi}$ of the single slit wave function sets the envelope of the diffraction pattern [see Eq.~(\ref{envelope})], which in the absence of interactions depends on the nominal (geometrical) width of the slit $a$. We will show that as a first order approximation the effect of a potential is a reduction of the effective slit width; this has been observed in various situations \cite{vdw}, a notable example being the C$_{60}$ experiments \cite{arndt}. We find:
\begin{equation}
\hat{\psi}(x) = \int_{\rm slit} \psi(\xi)\, \exp\left(\ii \frac{ 2\pi \xi x}{ \lambda L} \right)\, \d \xi 
\end{equation}
where $x$ denotes the transversal coordinate on the screen plane, consistently with our notation. As suggested by Ref.~\cite{vdw}, we approximate the above Fourier transform using a cumulants expansion \cite{abramowitz}  [a normalization of $\psi(\xi)$ ensuring $\int_{\rm slit} \psi(\xi)\, \d\xi =1$ is implied]
\begin{equation}
\log \hat{\psi}(x) = \sum_{n=1}^\infty \k_n \frac{(\ii \zeta)^n}{n!} \quad \mathrm{with} \quad
\zeta=\frac{2 \pi x}{\lambda L}
\end{equation}
where the cumulants $\k_n$ are defined in terms of the raw moments $\mu_k$ of $\psi(\xi)$, namely,
\begin{equation*}
\mu_k = \int \psi(\xi)\, \e^{\ii \varphi(\xi)} \,\xi^k\, \d\xi,
\end{equation*}
and we are interested in the first two terms only
\begin{equation}
\k_1 = \mu_1, \quad \mbox{and} \quad 
\k_2= \mu_2 - \mu_1^2 ,
\end{equation}
which, truncating the series to the second order, result in the approximation
\begin{eqnarray}
\hat{\psi}(x) &\approx \exp \left( \ii \k_1 \zeta - \k_2 \frac{\zeta^2}{2} \right)\\[1ex]
|\hat{\psi}(x)|^2 &  \approx \exp \left\{ -2 \zeta \, \Im{\rm m}[\k_1] - \Re{\rm e}[\k_2] \zeta^2 \right\}
\end{eqnarray}
In order to give a physical meaning to the expansion parameters, we analyze the case of no interaction, with a rectangular wave function $\psi(\xi) = a_0^{-1}\, \chi_{[-a_0/2 ,a_0/2]}$. The first raw moment (the mean) of this distribution vanishes due to parity $\psi(\xi) = \psi(-\xi)$, and in general we expect this to be true in any realistic situation, as both a reasonable interaction potential and the wall geometry will be symmetric with respect to $\xi \to -\xi$. The second moment is simply evaluated
\begin{equation*}
\k_2 = \mu_2 = \frac{1}{a_0} \int_{-a_0/2}^{a_0/2} \xi^2\, \d \xi = \frac{a_0^2}{12}
\end{equation*}
leading to:
\begin{equation*}
|\hat{\psi}(x)|^2 \approx \exp \left[ -\frac{1}{3} \left(\frac{\pi a_0 x}{\lambda L} \right)^2\right].
\end{equation*}
As one may expect, the exact expression is the well known $\hat{\psi}(x) = \mbox{sinc}[\pi a x/(\lambda L)]$, the central peak of the sinc function coincides with the Gaussian approximation, suggesting to identify an effective slit width as follows:
\begin{equation}
a_{\rm{eff}} = \sqrt{12\, \Re{\rm e}[\k_2]}.
\label{eqn:aeff}
\end{equation}
We will now distinguish the two cases of neutral and charged particles in term of the potential.

\subsection{Neutral particles}

The nonretarded van der Waals atom-surface potential \cite{vdw}, which affects all types of neutral polarizable particles, is expressed in terms of the distance from the surface and the coefficient $C_3$ as 
\begin{equation}
V_{\rm vdW}(r) = -\frac{C_3}{r^3}, \qquad (r>10 \unit{\mbox{\AA}}).
\label{eqn:vdwpot}
\end{equation}

It has been shown with direct electron microscope imaging, that diffraction gratings of the type commonly used for matter-wave experiments can have a trapezoidal (see Fig.~\ref{fig:wedgeslit}) slit profile \cite{vdw,savas,cornu,savas2,reticoli}, as a result of the fabrication process. Therefore it is useful to study this immediate generalization, from which the trivial parallel-plane profile is recovered in the limit $\beta \to 0$. Calculations also show that the introduction of even a small wedge angle has a significant impact on the effective width of the slits compared to parallel-planes approximation at the same thickness and interaction strength.

The potential can be written in terms of the distance of the generic point $(\xi,y)$ from the right and left grating walls, respectively $d_1$ and $d_2$. The symmetry of the system implies that these quantities are related by $d_2(\xi,y) = d_1(-\xi,y)$. Since $d_1(\xi,y) = a_m/2 - \xi + y\, \tan \beta$, as it is evident from Fig.~\ref{fig:wedgeslit}, the projection of the distance on the normal vector to the side wall is simply obtained by multiplying $d_2(\xi,y)$ for the cosine of the wedge angle $\beta$. The integration in Eq.~\eqref{eqn:vdwphaseshift} is straightforward, and yields for the phase shift of the potential \eqref{eqn:vdwpot}
\begin{equation}
\varphi_1(\xi,\beta) = \frac{C_3}{v}\, \left. \frac{ 2\left( \hbar \cos^3 \beta \tan \beta \right)^{-1}}{\left[a_m - 2(\xi - y \tan \beta) \right]^2} \right|_{y=0}^{y=\delta}.
\end{equation}
Taking both surfaces into account we obtain:
\begin{equation}
\varphi(\xi,\beta) = \varphi_1(\xi,\beta) + \varphi_1(-\xi,\beta).
\label{eqn:totalphaseshift}
\end{equation}
We note that at fixed geometry the ratio $R=C_3/v$ sets the overall scale of the interaction strength, as $\varphi(\xi,\beta) \propto C_3/v$. Numerical estimates of Eq.~\eqref{eqn:aeff} with the phase shift (\ref{eqn:totalphaseshift}) are in good agreement with the experimental and theoretical results in \cite{vdw}. Therefore, this kind of interactions can be accounted for by a reduction in effective slit width, at least up to the highest interaction strength tested experimentally, that is $R^{\rm max} = C_3/v \approx 2.74 \cdot 10^{-12} \unit{meV \cdot nm^2}$ (this is obtained for Kr atoms at $v=400 \unit{m/s}$ with a SiN$_x$ grating, and a measured $C_3=1.1 \unit{meV \cdot nm^3}$). Hoinkes' empirical rule \cite{hoinkes}, which has been confirmed experimentally \cite{vdw}, states that for a given material $C_3$ is linear in the particle static polarizability $\alpha$. The static polarizability of 	\positronium atoms, $\alpha^{\rm (Ps)}$, in the ground state (estimated treating it as an hydrogen-like atom with the appropriate reduced mass) is $\alpha^{\rm (Ps)} \equiv 8 \alpha^{\rm (H)} \approx 5.33 \unit{\mbox{\AA}^3}$, roughly twice that of the Kr atoms. In turn, Ps atoms have an interaction scale $R < R^{\rm{max}}$ down to speeds of $v_{\rm{min}}^{(\rm{Ps})} \approx 800 \unit{m/s}$, corresponding to a very low energy $E \approx 3.6 \cdot 10^{-3} \unit{meV}$, while the lower speed limit for an experiment with antihydrogen atoms (assuming $\alpha^{(H)}=\alpha^{(\overline{H})}$) would be $v_{\rm{min}}^{(\rm{\overline{H}})} \approx 100 \unit{m/s}$. Therefore, we can safely conclude that  a treatment of the van der Waals interaction in terms of Eq.~\eqref{eqn:aeff} is fully adequate to describe experiments involving current Ps sources \cite{positronium,positronium:2}.

\subsection{Charged particles}

The above procedure is easily generalized to all potentials depending on the distance from the surface as $ \propto r^{-n}$. Relevant examples are the retarded van der Waals interaction ($n=4$) and the electrostatic potential ($n=1$). In the AEgIS experiment at CERN \cite{aegis2013}, the production of antihydrogen and \positronium atoms also involves, as an intermediate step, the realization of a steady beam of charged antimatter, specifically $e^+$ and $\overline{p}$. It will be interesting to carry out interferometry experiments on these objects as well, because no successful demonstration of interference has been obtained for these systems, yet. Moreover, to the best of our knowledge, this statement also applies to any kind of charged system heavier than an electron.
\par
It is a standard result in electrostatics that the potential acting on a point charge $q$ sitting at an orthogonal distance $r$ from a dielectric surface (relative permittivity $\epsilon$) is that of a point charge $q' = q \left( 1 - \epsilon \right) \left(1+ \epsilon \right)^{-1}$ located on the axis of symmetry with respect to the plane of the surface. Therefore, for the same geometry represented in Fig.~\ref{fig:wedgeslit}, we can argue that the potential will be given by 
\begin{equation}
V_{\rm{el}}(r) =  \frac{1 - \epsilon}{1+ \epsilon} \frac{q^2}{4 \pi \epsilon_0} \frac{1}{2r}
\end{equation}
where $r$ the distance from the grating wall. By using Eq.~(\ref{eqn:vdwphaseshift}) we find:
\begin{equation}
\varphi_1^{\rm{el}}(\xi,\beta) = \frac{q^2 (1 - \epsilon) (1 + \epsilon)^{-1}}{8 \pi \epsilon_0 \hbar v \sin \beta} \log \left[ \frac{a_m - 2( \xi - \delta \tan \beta)}{a_m - 2\xi } \right],
\label{eqn:phaseshiftelectro}
\end{equation}
where $a_m$, $\delta$ and $\beta$ are the same as in Fig.~\ref{fig:wedgeslit}.
Once the total phase shift $\varphi(\xi,\beta) = \varphi_1^{\rm{el}}(\xi,\beta) + \varphi_1^{\rm{el}}(-\xi,\beta)$ is obtained, as in Eq.~\eqref{eqn:totalphaseshift}, we can follow the same procedure and account for the electrostatic interaction introducing an effective slit width, given by \eqref{eqn:aeff}. However, the orders of magnitude involved might be very different. If we compare the potential strengths at the center of a slit of width $a = a_m = a_M$ (for the sake of simplicity we set $\beta=0$), then
we obtain the following expression for the ratio:
\begin{equation*}
\frac{V_{\rm{el}}}{V_{\rm{vdW}}} = \frac{q^2}{4 \pi \epsilon_0} \frac{1 - \epsilon}{1 + \epsilon} \frac{a^2}{C_3},
\end{equation*}
which shows a quadratic dependence on the slit width, descending from the different power-law scaling of the potentials. Assuming that $q=e$, the electron charge, we also have
$e^2(4\pi \epsilon_0)^{-1} = 1439.964~\mbox{meV} \cdot \mbox{nm}$.
Consistently, we recall from the previous section, that $C_3$ is of the order of a few $\mbox{meV} \cdot \mbox{nm}^3$, in turn we have $V_{\rm{el}}/V_{\rm{vdW}} \sim  a^2\, 10^3 \mbox{[nm]}$.
Though the difference seems very large, this is only a pointwise estimate, while the effective slit width is determined by the behavior of the phase shift over the whole range of $\xi$. Nevertheless, we can see that there are realistic situations where the calculated impact of electrostatic interactions is indeed very high, as discussed in table \ref{tab:electro}. We see that in view of the higher typical potential strength in comparison with the van der Waals interaction, for low energy antiprotons Eq.~\eqref{eqn:aeff} predicts sizeable reduction in effective slit width, i.e., $\gtrsim 50 \% $. It turns out that probably the interaction is too strong in this regime to be treated with this approximation.
\begin{table}[htbp]
\centering
\begin{tabular}{|c|c|c|}
\hline 
Energy $[\unit{keV}]$ &  $a_{\rm{eff}} \quad e^+ \quad [\unit{nm}]$  & $a_{\rm{eff}} \quad \overline{p} \quad [\unit{nm}]$  \\ 
\hline 
0.1 & 401.3 & 148.1 \\ 
\hline 
1 & 477.2 & 285.8 \\ 
\hline 
10 & 497.1 & 397.4 \\ 
\hline 
100 & 499.7 & 460.0 \\ 
\hline 
\end{tabular} 
\caption{Calculated effective width for realistic parameters (order of magnitude) applicable to possible experiments, $a_0 = 0.5 \unit{\mu m}$, $\beta=5^\circ$, $\epsilon =4$, for \positron and \antiproton of varying energy. The grating thickness is set to $\delta = 500 \unit{nm}$ and $\delta = 160 \unit{nm}$ respectively, reflecting the typical scale necessary to absorb the particles completely outside the slits for silicon at $1 \unit{keV}$ reference energy. }
\label{tab:electro}
\end{table}
\par
Exact numerical calculation of the envelope function [still using the eikonal approximation for the phase shift (\ref{eqn:vdwphaseshift})] shows that for low energy particles there is indeed a stark departure from the \emph{sinc} shape expected for weak interactions.
Experimental data with electrons in the $0.5 \div 4 \unit{keV}$ energy range exist \cite{savas}: they refer to the same grating geometry described here, where although no explicit information on the effective slit width is reported, a model for an envelope function is developed and an analysis of the plotted data seems compatible with our theoretical predictions based on (\ref{eqn:phaseshiftelectro}), which yield a small $|a_{\rm{eff}} - a_0|/a_0 \lesssim 10 \%$ for that energy range.

\section{Talbot-Lau interferometry} \label{s:talbot}

In section~\ref{s:advanced} we have discussed how the contrast of interference patterns is affected by the ratio between the coherence length $l_0$ and the grating period $D$. This imposes technical constraints on the design of an interferometer using the geometry shown in Fig.~\ref{fig:gratingwithsource}. First of all the finite resolution of the detector has to be taken into account. This parameter greatly depends on the kind of detector and on the particles involved. For example, in the case of anti-hydrogen, $e^{+}$ and $\overline{p}$ emulsion detectors could be employed, which are capable of a spatial accuracy up to $0.6 \div 1 \unit{\mu m}$  \cite{aegis2013,emulsioni}.
Being $L$ the grating to detector distance and $D$ the grating period, the Fraunhofer diffraction orders are separated by
\begin{equation*}
\Delta x = \frac{L \lambda}{D}.
\end{equation*}
If $\delta x$ is the experimental sensitivity, in order to resolve each maximum of the diffraction pattern within at least an interval $M\, \delta x$, with $M$ integer, we should have
\begin{equation*}
\frac{L \lambda}{D} \geq M\, \delta x \Rightarrow L \geq M D\, \frac{\delta x}{\lambda},
\end{equation*}
which imposes a constraint on $L$. It is clear that for a fixed wavelength and geometry both increasing  $M$ and reducing the distance $L$, which is of utmost importance with decaying particles, requires a decrease in the grating period $D$.
\par
\begin{figure}
\begin{center}
\includegraphics[width=0.6\textwidth]{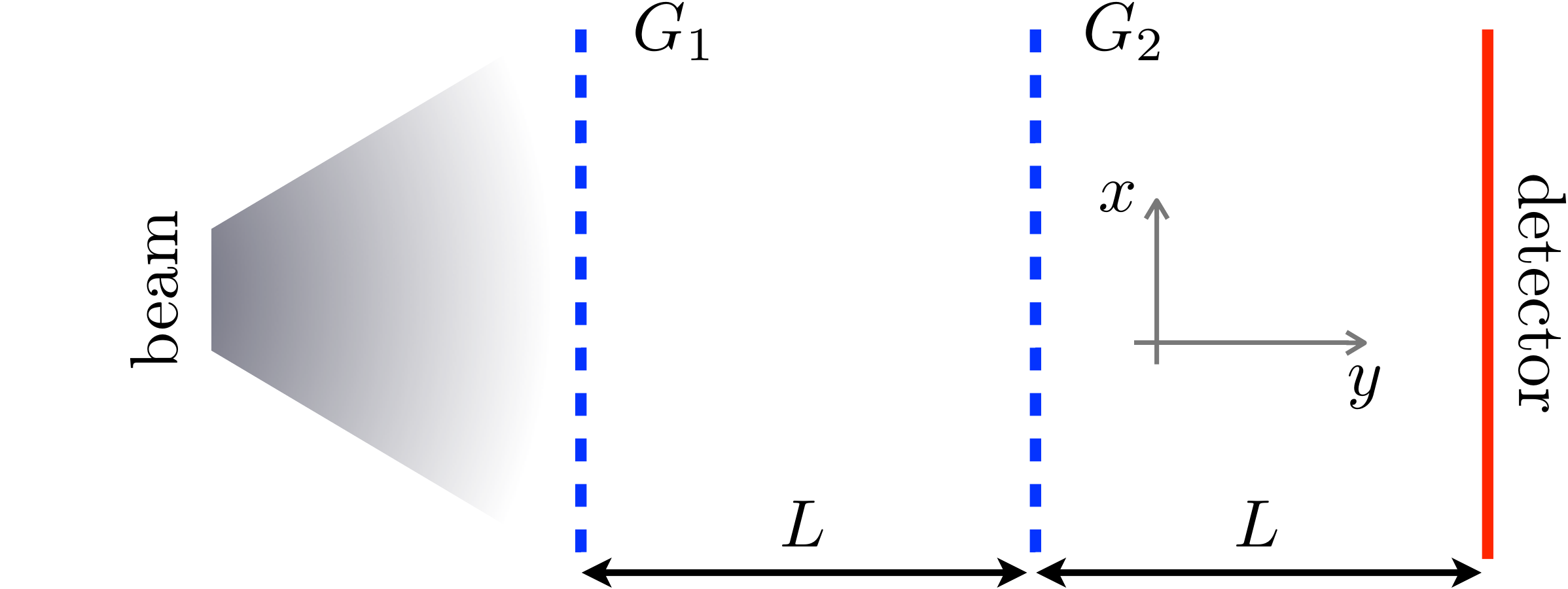}
\end{center}
    \vspace{-0.2cm}
  \caption{Sketch of a Talbot-Lau interferometer. The first grating ($G_1$) is illuminated by an incoherent beam of mean wavelength $\lambda$, and acts as an intensity mask providing the necessary coherence for illuminating the second grating ($G_2$). Their separation $L$ is set to the observation distance from $G_2$ and the gratings have the same period $D$ and slit width $a$. If $L$ matches the Talbot Length $T_L = D^2 / \lambda$, this setup produces on the detector plane high contrast fringes with period $D$. }
  \label{fig:threegrating}
\end{figure}

Moreover, starting from Eq.~\eqref{eqn:cohestimate} and requiring that the coherence length $l_0$ is at least $\tilde{M}$ times the grating spacing $D$, we obtain the following condition on the source-grating distance:
\begin{equation*}
y_s \geq  \tilde{M} D\, \frac{ 2\sigma_s}{\lambda}.
\end{equation*}
Therefore, to obtain a good coherence either the distance $y_s$ has to be increased or the source dimension $\sigma_s$ reduced as much as possible. Note that, apparently, by reducing the period $D$ we can satisfy both conditions on $L$ and $y_s$, however a reasonable small value for $D$ is fixed by the grating construction constraints. This poses technical challenges due to the particle decay in the first case (for ortho-Ps atoms with a lifetime $\tau=142 \unit{ns}$ \cite{pdg} and realistic thermal speeds $v \approx 10^5 \unit{m/s}$ \cite{positronium:3}, $y_s$ should be in the range of a few centimeters), and to difficulties in manipulating the beam size in the latter. 
In view of these consideration, we suggest that a different kind of interferometer would be best suited for experiments with antimatter, namely a Talbot-Lau setup \cite{clauseratominterferometry,patorski}, which is sketched in Fig.~\ref{fig:threegrating}.

It is worth noting that the usual Talbot-Lau configuration involves a third grating as a scanning mask \cite{clauseratominterferometry,clauserli}, that is not necessary in our case since we assume that the high resolution of the detector will allow to directly resolve the diffraction pattern. The fundamental property of this geometry is that it produces high contrast fringes regardless of the coherence and spatial extension of the illuminating beam. There are two physical phenomena governing this apparatus: the Talbot self-imaging effect \cite{patorski}, stating that in the Fresnel region of a coherently illuminated periodic grating self-images of the grating transmission function will appear at $L=n T_L = n D^2 / \lambda$ (as well as rescaled sub-images with a fractional period for half-integer multiples ), and the so called Lau effect \cite{lau}. This effect can be understood as arising from an incoherent superposition of patterns produced by laterally displaced, mutually independent, point sources, the role of which in the apparatus of Fig.~\ref{fig:threegrating} is played by the slits of the first grating  \cite{patorski}. The periodic images thus produced can overlap ``constructively'' if the first grating has a suitable periodicity; under these conditions the elementary displacement on the screen plane produced by moving between adjacent sources equals the Talbot image period or an arbitrary integer multiple of the latter. In particular, this ``resonance'' condition is met in the configuration of Fig.~\ref{fig:threegrating} when $L=T_L$.
Geometrically, this setup bears a strong similarity to a classical \emph{moir\'e deflectometer} \cite{moiree}. What discriminates between the purely classical and the quantum interference regime is the condition for diffraction to be negligible, namely  \cite{moiree}:
\begin{equation}
\frac{L \lambda}{D} \ll D.
\label{eqn:diffcondition}
\end{equation}

\begin{figure}[tb]
  \begin{center}
    \includegraphics[width=0.8\textwidth]{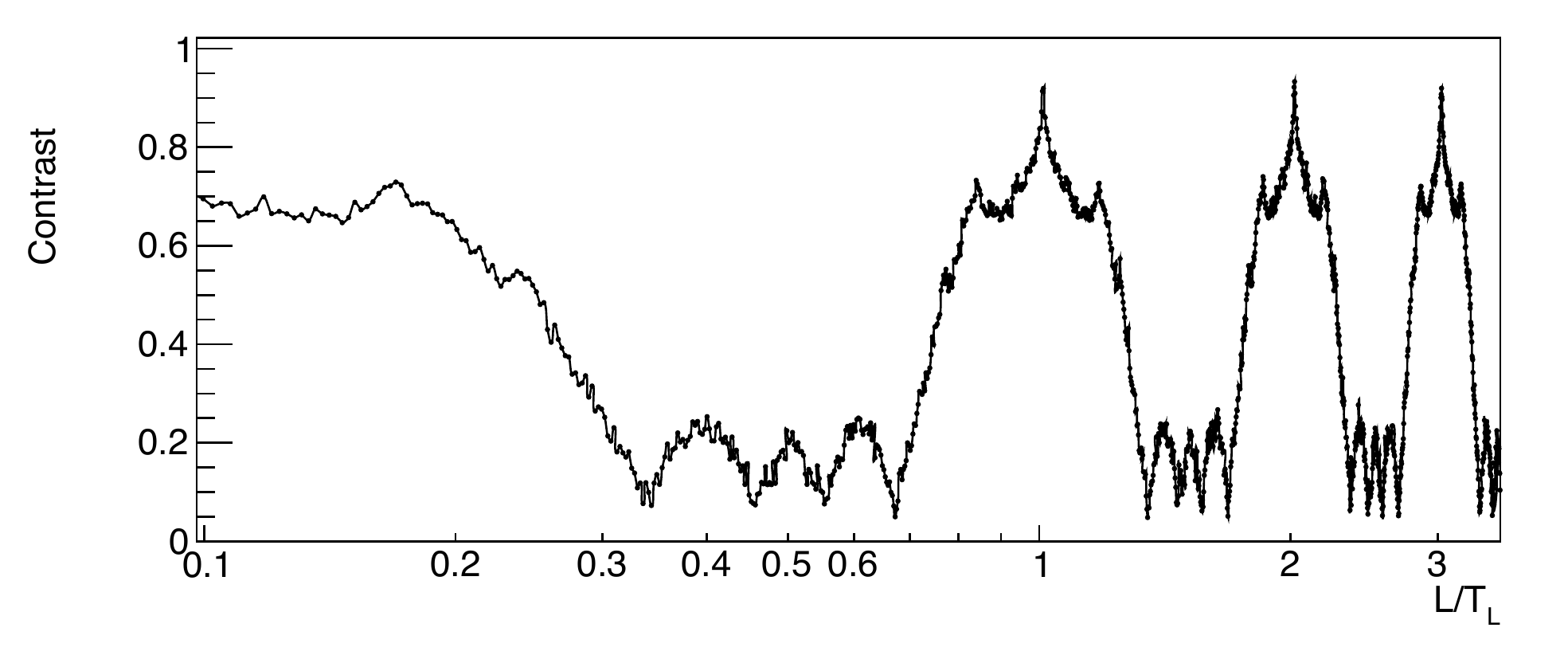}
    \vspace{-0.8cm}
  \end{center}
  \caption{Monte Carlo simulation of the fringe visibility modulation (contrast) as a function of $L/T_{L}$ (Log scale), for a well defined particle velocity. For definiteness the following parameters (realistic for an experiment with $1 \unit{keV}$ antiprotons) were chosen $D_1=D_2 =D = 265 \unit{nm}$, $N=40$ and $a=90 \unit{nm}$. The calculated Talbot length for this period and energy is $T_L = 77.9 \unit{mm}$. Note that the typical contrast peaks disappear when $L/T_L \ll 1$: this corresponds to a classical behavior of the particles.}
  \label{fig:visibilityspeed}
\end{figure}

A moir\`e deflectometer and a Talbot-Lau interferometer as defined in Fig.~\ref{fig:threegrating} have in common that they produce a fringe pattern with period $D$. The question that now naturally arises is: how can the experimental results prove that the observed fringes are a true interference effect and not simple classical geometrical shadow patterns produced by ballistic particles? As mentioned, high contrast fringes are expected only if the grating separation is an integer multiple of the Talbot length, while for ``classical projectiles'' the contrast does not depend on this condition. This property ultimately descends from the longitudinal periodicity of the so-called \emph{Talbot carpet}, which is a distinctive feature of diffraction in the Fresnel region. Therefore, the observation of this kind of pattern is a proof of the wave character of the interfering particles. From the experimental point of view, this can be done by continuously adjusting the grating separation or changing the particle energy (hence the Talbot Length) in a monochromatic beam, and measuring the modulation in contrast as a function of the parameter $L/T_L$. If the apparatus is truly operating as an interferometer and not as a classical device, distinct peaks in contrast should be detected, as shown in Fig.~\ref{fig:visibilityspeed} \cite{sala:thesis}. Recalling \eqref{eqn:diffcondition}, we see that the classical limit corresponds to $L \ll T_L$, as it is confirmed by numerical calculations showing a weak dependence of the contrast on $L$ in this region. This is clear from Fig.~\ref{fig:visibilityspeed}: as the grating distance $L$ falls below $T_L = 77.9$~mm (given by the simulated period and energy) the contrast peaks disappear thus revealing a classical behavior of the particles. \STErev{We note that in the simulation of Fig.~\ref{fig:visibilityspeed} we assumed an incoherent particle beam, thus we applied Eq.~(\ref{I:bar}) treating the slits of the first grating as a collection of extended incoherent sources.}

Another advantage of Talbot-Lau configuration is its robustness with respect to external stray fields.
Suppose for example that the interferometer is subjected to uniform external electric and magnetic fields which exerts a force $F$ on the charged particles: this force will be negligible if the corresponding deviation from a straight trajectory is smaller than the typical size of the finest structure in the observable pattern. Let us call this quantity $\Delta$, and introduce the flight time of the particle $\tau = L/v$. Using the identity $\Delta = F_{\rm{crit}}\tau^2/m   $, we obtain the following relation:
\begin{eqnarray*}
F_{\rm{crit}}
= \frac{ h^2 D}{m L^2\lambda^2 },
\end{eqnarray*}
where for a Talbot-Lau geometry $\Delta=D$. If we set $L = T_L = D^2/\lambda$, we have:
\begin{eqnarray}
F_{\rm{crit}} &= \frac{h^2 }{m D^3} ,
\label{eqn:critfield}
\end{eqnarray}
that is independent of the particle energy. Furthermore, from the Lorentz force $F = q(E + vB)$ one can deduce the critical values of the involved electric and magnetic fields by the simple relations $E_{\rm{crit}} = F_{\rm{crit}} /q $ and $B_{\rm{crit}} = F_{\rm{crit}} /qv$. It is worth noting that in the case of Fraunhofer interferometry (far field) a similar calculation with $\Delta=L\lambda/D $ leads to a critical force $F_{\rm{crit}}^{(F)} = h^2 / (m L D \lambda) \propto v$ which is thus affected by the particle energy.
\par
Overall, the Talbot-Lau geometry has several advantages which are especially relevant for anti-matter interferometry. In particular, it allows to minimize the total length of the apparatus, a crucial requirement for decaying Ps, and to employ a larger source with weak coherence requirements, significantly increasing the particle flux.

\subsection{Realistic application}

We now present a realistic estimate of the expected contrast signal in an $e^+$ experiment, following the experimental methodology outlined in this section (see Figs.~\ref{fig:threegrating} and \ref{fig:visibilityspeed}). We assume that the particle mean energy $E_0$ can be tuned between $5 \unit{keV}$ and $ 20 \unit{keV}$ (reasonable for current continuous $e^+$ beams) with a narrow Gaussian energy distribution ($\sigma / E_0 \lesssim 2 \%$). By varying the energy we scan the ratio $L/T_L$, where the Talbot length reads:
\begin{equation}
T_L= \frac{D^2 \sqrt{2 m E_0}}{h}.
\label{eqn:rapportoproposal}
\end{equation}
A contrast peak is expected around $L \approx T_L$. The above formula makes it clear that to scan this region there are specific complications related to each choice of which parameter to vary: the energy is bounded by technical constraints and furthermore provides a sub-linear scaling, whereas the grating distance can be varied arbitrarily. There are however technical complications in physically moving an apparatus sensitive to alignment over considerable lengths.

We recall that in this configuration $D$ sets the periodicity of the interference pattern, which should be larger than the detector resolution. However as evident from (\ref{eqn:rapportoproposal}), the Talbot length scales quadratically with $D$. Thus it rapidly becomes very large, and too long an apparatus poses additional challenges related both to grating alignment and shielding of a larger region from stray fields. We set $D=2 \unit{\mu m}$ corresponding to a Talbot Length $T_L = 0.326 \unit{m}$ at the median energy of the considered energy range, namely $E_0=10 \unit{keV}$. Using Eq.~(\ref{eqn:critfield}) we obtain $E_{\rm{crit}} \approx 0.2 \unit{V/m}$ and $B_{\rm{crit}} \approx 0.3 \unit{mG}$ for the maximum tolerable electric and magnetic field, respectively, evaluated at a $E_0=5 \unit{keV}$, which corresponds to the worst scenario.

The maximum magnetic field is particularly critical, as the requirement is smaller than the natural magnetic field of the Earth, however considering that experiments with electrons and similar length scales involved have been successfully carried out \cite{threeelectron}, we believe that an appropriate mu-metal based shielding will be enough to circumvent the problem. Moreover an uniform constant magnetic field will only rigidly shift the pattern in space \cite{arndtonatomint}, in fact the above limits indicate the maximum allowed \emph{fluctuations} (either in time or in space) of the $E$ and $B$. 

Another problem our theoretical analysis allows to account for is the electrostatic interaction with the grating walls. First of all we have to set the slit width, and thus the open fraction of the grating; this is best set at $a/D \approx 30 \%$, implying $a=0.6 \unit{\mu m}$. Higher values could improve the total particle flux minimizing the losses inside the material grating, but will also reduce the contrast due the overlapping diffraction peaks.
We remark that for the gratings to work as true intensity masks, their thickness must be sufficient to stop all the positrons outside the slits. For such low energies few microns of SiN$_x$ will be sufficient.
Therefore, assuming these parameters and a small wedge angle $\beta = 10 ^{\circ}$, a grating thickness $\delta= 800 \unit{nm}$ and $E_0=5 \unit{keV}$, equation (\ref{eqn:aeff}) predicts an effective slit width $a_{\rm{eff}} = 0.598 \unit{\mu m}$. This deviation is very small ($\approx 0.5 \%$), thus the effect will be completely negligible for the considered choice of parameters.

\begin{figure}
  \centering
  \includegraphics[width=0.4\textwidth]{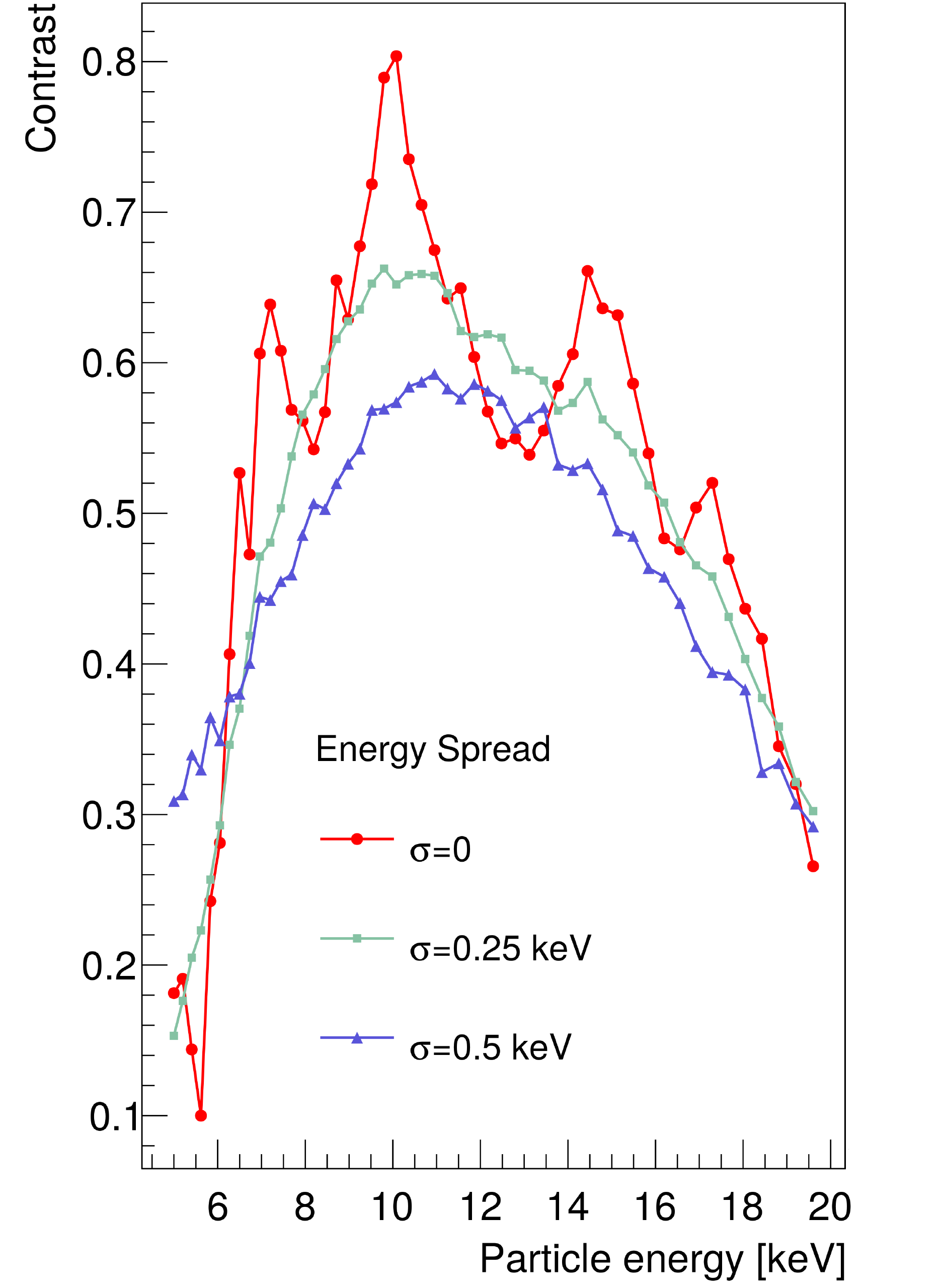}
  \vspace{-0.3cm}
  \caption{Monte Carlo simulations of the expected contrast modulation as a function of $E_0$, for the monochromatic case and with a Gaussian distribution of increasing $\sigma$ centered on $E_0$. The total number of slits is set to $N=40$, sufficiently large to provide realistic results. See text for more details on the parameters.}
  \label{fig:proposalconfronto}
\end{figure}

Having evaluated the most relevant effects and set the geometry of the apparatus, we can now use Eq.~(\ref{I:bar}) to predict the measured contrast modulation \cite{sala:thesis}. The result is shown in Fig.~\ref{fig:proposalconfronto}, which contains interesting indications. On the one hand we can see that a particle energy between $5 \unit{keV}$ and $20 \unit{keV}$ is enough to observe a full peak. On the other hand, we see that also a broader energy distribution ($\sigma = 0.25$~keV or $\sigma = 0.5$~keV in Fig.~\ref{fig:proposalconfronto}) still allows for a good visibility of the contrast modulation.

\section{Conclusion and outlook}\label{s:conclusion}

In this paper we reviewed the basic elements of diffraction theory applied to matter-wave interferometry. In particular, we focus our analysis on the possible issues arising from the use of charged or neutral antimatter particles. To sustain our investigation we also performed Monte Carlo simulated experiments based on realistic parameters. In particular, we have considered the effect due to a realistic source (extended and non-monochromatic) and the interaction with the grating as well as the influence of stray electromagnetic fields. \STErev{We also found that van der Waals interactions with the material grating become critical for highly polarizable particle systems. In this scenario a possible solution could be resorting to light gratings \cite{lg:02,lg:09}.} We have shown that the better configuration to carry out matter-wave interferometry with decaying particles is given by the Talbot-Lau setup also in the presence of a Gaussian distribution of the particle energy, which realistically describes the actual $e^{+}$ and $\overline{p}$ beams. Furthermore, exploiting the high resolution capabilities of the antiparticle detectors, such as the nuclear emulsions, we have shown that the typical Talbot-Lau setup involving three gratings can be reduced to a two-grating configuration which indeed simplifies the experimental implementation. Our analysis paves the way to further investigations in order to design an experiment to demonstrate antimatter-wave interference \STErev{also in view of possible applications in the emerging field of gravity experiments using antimatter \cite{lg:02,ham:14,agh:nat}}.

\section*{Acknwledgments}
The authors would like to thank R.~Ferragut and C.~Pistillo for useful discussions and suggestions. SO acknowledges financial  support by MIUR (project FIRB ``LiCHIS'' - RBFR10YQ3H) and by EU through the Collaborative Project QuProCS (Grant Agreement 641277).

\section*{References}

\bibliographystyle{iopart-num}

\providecommand{\newblock}{}

\end{document}